\documentclass[subeqn,11pt,a4paper]{article}
\usepackage{graphicx} 
\usepackage{amssymb,amsmath,amsfonts,amsthm, amscd, mathrsfs, helvet, mathtools}
\usepackage{hyperref}
\usepackage{tikz-cd}

\theoremstyle{plain}

\newtheorem*{theorem*}{Theorem}

\newtheorem*{proposition*}{Proposition}

\theoremstyle{definition}

\def\CC{\mathbb{C}}

\def\RR{\mathbb{R}}

\newcommand{\g}{\mathfrak{g}}

\numberwithin{equation}{section}

\begin{document}

\title{\textbf{On the geometry of Lagrangian one-forms\\}\vspace{8pt}}

\author{Vincent Caudrelier\footnote{Corresponding author: v.caudrelier@leeds.ac.uk} \and Derek Harland}
\date{%
{School of Mathematics, University of Leeds, Leeds LS2 9JT, United Kingdom}\\%
}

\maketitle

\begin{abstract}
Lagrangian multiform theory is a variational framework for integrable systems.  In this article we introduce a new formulation which is based on symplectic geometry and which treats position, momentum and time coordinates of a finite-dimensional integrable hierarchy on an equal footing.  This formulation allows a streamlined one-step derivation of both the multi-time Euler-Lagrange equations and the closure relation (encoding integrability).  We argue that any Lagrangian one-form for a finite-dimensional system can be recast in our new framework.
This framework easily extends to non-commuting flows and we show that the equations characterising (infinitesimal) Hamiltonian Lie group actions are variational in character.  We reinterpret these equations as a system of compatible non autonomous Hamiltonian equations.
\end{abstract}


\vfill


\newpage

\section{Introduction}
The geometry of integrable systems has been dominated by the Hamiltonian formalism, symplectic and Poisson geometry, with the celebrated Liouville--Arnold theorem as the cornerstone of this edifice. Comparatively very recently, much work has been devoted to defining and describing integrability within a purely variational framework, as an effort to restore the natural balance between Hamiltonian and Lagrangian formalisms in the realm of integrable systems. Pioneered in \cite{LN}, Lagrangian multiform theory offers a variational framework to describe and study (classical) integrability by harnessing the concept of integrable {\it hierarchies}. It applies equally well in discrete and continuous finite-dimensional ($d=1$) integrable systems \cite{YKLN,Su1,PS,CDS,CSV}, discrete \cite{LN,XNL,RV,NZ} and continuous \cite{SuV,SNC1,SNC2,CS1,CS2,PV,CSV2} integrable field theories in $1+1$ dimensions ($d=2$), discrete \cite{LN2,Nij2} and continuous \cite{SNC3,Nij} field theories in $2+1$ dimensions ($d=3$) and even in semi-discrete models \cite{SV,Nij2}.

The underlying ideas are the same in all contexts: one considers a differential (difference) $d$-form on an $N>d$ space called a multi-time and forms an action by integrating (summing) this form over an arbitrary $d$-dimensional submanifold (sublattice) of the multi-time. A generalised variational principle is then applied to derive the structural equations of the theory: 1) the multi-time Euler-Lagrange equations obtained by varying over the degrees of freedom for an arbitrary choice of the submanifold (sublattice) and requiring criticality of the action; 2) the {\it closure relation} obtained by varying the underlying submanifold (sublattice) and requiring criticality of the action {\it on-shell} (on solutions of the multi-time Euler-Lagrange equations). Crucially, the closure relation is the variational equivalent \cite{Su1,CDS} of the well-known Poisson involutivity of Hamiltonians defining Liouville integrability.

In the standard approach to multiform theory described above, dependent and independent variables are varied separately in steps 1) and 2).  In this sense, it is an unsatisfactory realisation of the original paradigm, which intended to put dependent and independent variables on an equal footing.  In Section \ref{symplectic_formulation} of this article we introduce a new framework for Lagrangian one-forms in which dependent and independent variables are varied simultaneously.  This results in a simpler derivation of the variational equations.

Our new framework is formulated in phase space.  This makes it is straightforward to write down a Lagrangian one-form for any finite-dimensional Liouville integrable system.  Previously, position-space Lagrangian one-forms were constructed by ad hoc methods and it was not known whether every integrable system admits a Lagrangian one-form.  We will show in section \ref{sec:equivalence} how to write down a position-space Lagrangian one-form for any Liouville integrable system by applying a generalised Legendre transform to the phase-space Lagrangian one-form.

Our new framework is rooted in symplectic geometry and as a result is very flexible.  In particular, we show in section \ref{sec:form_group} that abelian multi-time can be replaced by a nonabelian (connected) Lie group, an idea first introduced in \cite{CNSV}.  The resulting variational principle delivers a Hamiltonian action of the Lie group.  To our knowledge, this is the first time that the notion of Hamiltonian group action is {\it derived from a variational principle}. The moment map of this action plays a very natural role in our Lagrangian one-form. It arises as the on-shell evaluation of a map which is the analog of the potential term in a traditional (Newtonian) Lagrangian. To summarise, our main results are:

\begin{enumerate}

\item The equivalent reformulation of the previous two-step variational principle of Lagrangian multiform theory into a single variational principle applied to an action for a {\it phase-space} Lagrangian one-form;

\item A systematic method to construct a Lagrangian one-form directly from the Hamiltonians and the symplectic form of a Liouville integrable system;

\item A variational formulation of Hamiltonian Lie group actions.
\end{enumerate}

The paper is organised as follows. In Section \ref{sec:2pbs}, we review the known generalised variational principle for Lagrangian one-forms in a form suitable for our purposes. This serves to introduce the objects and notations. Then, we introduce our reformulation based on a phase-space Lagrangian one-form. In Section \ref{sec:equivalence}, we prove the equivalence of the two pictures, give a simple illustration with the Toda chain, and discuss the case where the Lagrangian one-form is linear in the velocities. The latter point is motivated by the fact that Lagrangian one-forms constructed in \cite{CDS,CSV} for large classes of integrable models are of this type. 
In Section \ref{sec:form_group}, we present a generalisation of the univariational principle to the case where multi-time $\RR^n$ is replaced by a Lie group. Applying this to a natural Lagrangian one-form, we establish for the first time that the equations describing (infinitesimal) Hamiltonian group actions of a (connected) Lie group on a symplectic manifold are variational: they appear as the Euler-Lagrange equations of our univariational principle.
We also give an interpretation of the equations as describing a collection of compatible {\it non autonomous} Hamiltonian equations. Section \ref{sec:ccl} contains our conclusions and some perspective on future directions. 

\section{Two variational problems}\label{sec:2pbs}
\subsection{Position-space Lagrangian one-forms}\label{traditional_formulation}
The traditional perspective on Lagrangian one-forms is as follows (we reformulate the derivation and results in \cite{Su1}, see also \cite{PPY}).  Let $q^\mu$ be $m$ real functions of $n$ real variables $t^j$.  Let $\Gamma$ be a curve in $\RR^n$, written parametrically as $s\mapsto t^j(s)$ for $s\in[0,1]$.  We introduce an action
\begin{equation}\label{action traditional}
    S[q,\Gamma]=\int_0^1 L_k[q^\mu,q^\mu_j]\frac{dt^k}{ds}ds.
\end{equation}
Here $q^\mu_j=\partial q^\mu/\partial t^j$.  In this equation and throughout the article repeated indices are implicitly summed over, unless stated otherwise. We will also always assume that boundary conditions are chosen so that boundary terms drop out when performing integration by parts. The integrand $L_k dt^k$ is referred to as a Lagrangian one-form and $L_k$ are the Lagrangian coefficients. The action $S$ depends on the map $q:\RR^n\to\RR^m$ and the parametrised curve $\Gamma\subset\RR^n$.
We seek a map $q$ such that for all curves $\Gamma$, $S[q,\Gamma]$ is critical with respect to variations of both $q$ and $\Gamma$.
We refer to this requirement as the {\it bivariational principle} (reflecting the fact that it involves two steps).
We now derive the associated variational equations.

First we consider variations $q^\mu\mapsto q^\mu+\delta q^\mu$.  The variation of $S$ is
\begin{equation}\label{q variation}
    \delta S = \int_0^1 \left(\delta q^\mu\frac{\partial L_k}{\partial q}+\delta q^\mu_j\frac{\partial L_k}{\partial q^\mu_j}\right)\frac{dt^k}{ds}ds,
\end{equation}
in which $\delta q^\mu_j=\partial \delta q^\mu/\partial t^j$.  Suppose that the variation $\delta q^\mu$ vanishes along the curve $\Gamma$.  Then
\begin{equation}\label{q constraint}
    \delta q^\mu(t^j(s))=0\quad \text{and}\quad 0=\frac{d}{ds}\delta q^\mu = \frac{dt^k}{ds}\delta q^\mu_k(t^j(s)).
\end{equation}
So for each $\mu$ the vector with components $\delta q^\mu_k$ is orthogonal to the vector with components $dt^k/ds$.  Then \eqref{q variation} vanishes for all $\delta q^\mu$ satisfying \eqref{q constraint} if and only if the components of $\frac{\partial L_k}{\partial q^\mu_j}\frac{dt^k}{ds}$ orthogonal to $dt^j/ds$ are zero, in other words,
\begin{equation}\label{equal momenta}
    \frac{\partial L_k}{\partial q^\mu_j}\frac{dt^k}{ds}=p_\mu\frac{dt^j}{ds}
\end{equation}
for some $p_\mu$.  This equation says that, for fixed $\mu$, the vector $dt^k/ds$ is an eigenvector of the matrix $M_\mu$ with entries $\partial L_k/\partial q^\mu_j$.  Our variational principle demands that this is true for all curves $t^k(s)$, and hence for all vectors $dt^k/ds$.  So every vector in $\RR^n$ is an eigenvector of $M_\mu$ with eigenvalue $p_\mu$, and this matrix must equal $p_\mu$ times the identity.  Thus the bivariational principle requires that
\begin{equation}\label{EL1}
    \frac{\partial L_k}{\partial q^\mu_j}=p_\mu\delta^j_k.
\end{equation}
for some $p_\mu$.  The traceless part of \eqref{EL1} imposes constraints on $q^\mu,q^\mu_j$ and the trace part then determines $p_\mu$ as a function of $q^\mu,q^\mu_j$.  We note that this equation imposes constraints on the Lagrangian one-form: there are choices of $L_k dt^k$ for which it cannot be solved.  

Now we consider more general variations $\delta q^\mu$ that are not necessarily 0 along $\Gamma$.  Assuming that the variational equation \eqref{EL1} is satisfied, \eqref{q variation} gives
\begin{multline}
    \delta S=\int_0^1\left(\delta q^\mu\frac{\partial L_k}{\partial q^\mu}+\delta q^\mu_kp_\mu\right)\frac{dt^k}{ds}ds
    =\int_0^1\left(\delta q^\mu\frac{\partial L_k}{\partial q^\mu}\frac{dt^k}{ds}+p_\mu\frac{d\delta q^\mu}{ds}\right)ds\\
    =\int_0^1\delta q^\mu\left(\frac{\partial L_k}{\partial q^\mu}\frac{dt^k}{ds}-\frac{d p_\mu}{ds}\right)ds
    =\int_0^1\delta q^\mu\left(\frac{\partial L_k}{\partial q^\mu}-\frac{\partial p_\mu}{\partial t^k}\right)\frac{dt^k}{ds}ds.
\end{multline}
This vanishes for all variations $\delta q^\mu$ and all curves $\Gamma$ if and only if
\begin{equation}\label{EL2}
    \frac{\partial L_k}{\partial q^\mu}-\frac{\partial p_\mu}{\partial t^k}=0
\end{equation}
Thus $S[q,\Gamma]$ is stable to variations of $q$ for all curves $\Gamma$ if and only if equations \eqref{EL1} and \eqref{EL2} hold.  In the literature on Lagrangian multiforms, these equations are known as multi-time Euler--Lagrange equations and more commonly expressed as
\begin{equation}
    \frac{\partial L_k}{\partial q^\mu_j}=0,\quad \frac{\partial L_k}{\partial q^\mu_k}=\frac{\partial L_j}{\partial q^\mu_j},\quad \frac{\partial L_k}{\partial q^\mu}-\frac{\partial}{\partial t^k}\frac{\partial L_k}{\partial q^\mu_k}=0
\end{equation}
for all $k\neq j$, in which there is no summation over repeated indices.

Having discussed variations of $q^\mu$, we consider variations of the curve $\Gamma$ of the form $t^k(s)\mapsto t^k(s)+\delta t^k(s)$.  The variation of $S$ is
\begin{equation}
    \delta S = \int_0^1 \left(\frac{\partial L_k}{\partial t^j}\delta t^j\frac{dt^k}{ds} + L_j\frac{d\delta t^j}{ds}\right) ds
    = \int_0^1 \left(\frac{\partial L_k}{\partial t^j}\frac{dt^k}{ds} -\frac{d t^k}{ds}\frac{\partial L_j}{\partial t_k}\right)\delta t^j ds.
\end{equation}
This vanishes for all variations $\delta t^j$ and all curves $t^j(s)$ if and only if
\begin{equation}\label{EL3}
    \frac{\partial L_k}{\partial t^j} -\frac{\partial L_j}{\partial t_k}=0.
\end{equation}
Equation \eqref{EL3} is known as the closure relation.
Note that in this equation $L_k$ depends on $t^j$ through $q^\mu(t_j)$, since we work with Lagrangian coefficients not depending explicitly on $t^j$; thus
\begin{equation}\label{djLk}
    \frac{\partial L_k}{\partial t^j}=\frac{\partial L_k}{\partial q^\mu}q^\mu_j+\frac{\partial L_k}{\partial q^\mu_i}q^\mu_{ij}.
\end{equation}
Equation \eqref{EL3} cannot be satisfied for all functions $q^\mu$; we only demand that it is satisfied for solutions $q^\mu$ of \eqref{EL1}, \eqref{EL2}.  In other words, we require that $L_k dt^k$ is closed ``on-shell''.

The closure relation \eqref{EL3} implies that the multi-time Euler-Lagrange equations \eqref{EL2} are generated by Poisson commuting Hamiltonians $H_i$.  These Hamiltonians are defined as usual:
\begin{equation}
    H_i := p_\mu q^\mu_i - L_i.
\end{equation}
A priori, these Hamiltonians are functions of $p_\nu$, $q^\mu$ and $q^\mu_j$, and we may compute their partial derivatives, treating these variables as independent:
\begin{align}
\frac{\partial H_i}{\partial p_\mu} &= q_i^\mu\label{eq_Ham1}\,,\\
\frac{\partial H_i}{\partial q^\mu} &= -\frac{\partial L_i}{\partial q^\mu}\,,\\
\frac{\partial H_i}{\partial q^\mu_j} &= p_\mu\delta_i^j-\frac{\partial L_i}{\partial q^\mu_j}\,.\label{eq_Ham3}
\end{align}
Equations \eqref{EL1} and \eqref{eq_Ham3} imply that the derivative with respect to $q^\mu_j$ is zero, so that $H$ may be regarded as a function of $p_\mu$ and $q^\mu$.
Equations \eqref{EL2} and \eqref{eq_Ham1} imply that the variables $p_\mu,q^\mu$ follow the Hamiltonian flows generated by $H_i$:
\begin{equation}
    \frac{\partial q^\mu}{\partial t^i}=\frac{\partial H_i}{\partial p_\mu},\quad \frac{\partial p_\mu}{\partial t^i}=-\frac{\partial H_i}{\partial q^\mu}.
\end{equation}
Moreover, equation \eqref{djLk} becomes
\begin{equation}
    \frac{\partial L_k}{\partial t^j}=\frac{\partial H_k}{\partial q^\mu}\frac{\partial H_j}{\partial p_\mu}+\frac{\partial L_k}{\partial q^\mu_i}q^\mu_{ij}+p_\mu q^\mu_{kj}.
\end{equation}
So the closure relation \eqref{EL3} is equivalent under the equations of motion \eqref{EL1}, \eqref{EL2} to
\begin{equation}
    0=-\frac{\partial H_k}{\partial q^\mu}\frac{\partial H_j}{\partial p_\mu}+\frac{\partial H_j}{\partial q^\mu}\frac{\partial H_k}{\partial p_\mu}=\{H_k,H_j\},
\end{equation}
where we introduced the canonical Poisson bracket $\{~,~\}$. Hence the existence of solutions to the bivariational problem requires that the Hamiltonians $H_i$ Poisson commute.

\subsection{Phase-space Lagrangian one-forms}\label{symplectic_formulation}

In this section, we introduce a different Lagrangian one-form and formulate our {\it univariational principle}. We then derive the corresponding univariational Euler--Lagrange equations. These encompass both the multi-time Euler-Lagrange equations {\it and} the closure relation of the traditional bivariational principle reviewed in the previous section. We show that existence of solutions to the univariational equations is equivalent to Poisson involutivity of the Hamiltonian functions.  Finally, we formulate the univariational principle in the setting of symplectic geometry.

Let $p_1,q^1,\ldots,p_m,q^m$ be coordinates on a $2m$-dimensional manifold $M$, let $H_1,\ldots H_n$ be $n$ real functions on $M$, let $t^1,\ldots,t^n$ be standard coordinates on $\RR^n$, and consider the one-form
\begin{equation}
\label{basis_one_form}
    L=p_\mu dq^\mu-H_i dt^i
\end{equation}
on $M\times \RR^n$.  For any parametrised curve $\gamma:[0,1]\to M\times \RR^n$ we define an action
\begin{equation}\label{geometric action}
    \mathcal{S}[\gamma]=\int_0^1\gamma^\ast L.
\end{equation}
More concretely, if we write $\gamma(s)=(p_\mu(s),q^\mu(s),t^i(s))$ then
\begin{equation}
    \mathcal{S}[\gamma]=\int_0^1\left(p_\mu\frac{d q^\mu}{ds}-H_i(p_\mu,q^\mu)\frac{dt^i}{ds}\right)ds.
\end{equation}
The action $\mathcal{S}$ is stable with respect to variations of $\gamma$ if and only if
\begin{equation}\label{VE}
    \gamma'\lrcorner dL = 0.
\end{equation}
Explicitly, with $\gamma'=\frac{d q^\mu}{ds}\frac{\partial}{\partial q^\mu}+\frac{d p_\mu}{ds}\frac{\partial}{\partial p_\mu}+\frac{d t^k}{ds}\frac{\partial}{\partial t^k}$, we obtain
\begin{align}
    \frac{dq^\mu}{ds}&=\frac{dt^i}{ds}\frac{\partial H_i}{\partial p_\mu}\,,\label{VE1}\\
    \frac{dp_\mu}{ds}&=-\frac{dt^i}{ds}\frac{\partial H_i}{\partial q^\mu}\,,\label{VE2}\\
    0&= \frac{d q^\mu}{ds}\frac{\partial H_i}{\partial q^\mu}+ \frac{d p_\mu}{ds}\frac{\partial H_i}{\partial p_\mu}\,.\label{VE3}
\end{align}
Note that if $t^i(s)=sv^i$ for some vector $v^i$ then the first two equations describe the Hamiltonian flow for $v^iH_i$.  

Now let $\Sigma\subset M\times\RR^n$ be an $n$-dimensional hypersurface.  We introduce a {\it univariational principle}, which demands that every curve $\gamma\subset\Sigma$ is a critical point of $\mathcal{S}[\gamma]$.  In other words, every $\gamma:[0,1]\to\Sigma$ solves \eqref{VE}. The term {\it univariational} is chosen because the principle involves one rather than two steps. To see what the principle means, we assume that $\Sigma$ is the graph of a function $\RR^n\to M$, in other words, $\Sigma$ is parametrised as $t^i\mapsto (p_\mu(t^i),q^\mu(t^i),t^i)$ (we will show later that this is not an assumption but is a consequence of the univariational principle).
Then $\Sigma$ solves the univariational principle if and only if
\begin{align}
    \frac{\partial q^\mu}{\partial t^i}&=\frac{\partial H_i}{\partial p_\mu}\label{usual_eqs1}\\
    \frac{\partial p_\mu}{\partial t^i}&=-\frac{\partial H_i}{\partial q^\mu}\label{usual_eqs2}\\
    0&= \frac{\partial  q^\mu}{\partial t^j}\frac{\partial H_i}{\partial q^\mu}+ \frac{\partial p_\mu}{\partial t^j}\frac{\partial H_i}{\partial p_\mu}\label{usual_eqs3}.
\end{align}
We refer to these equations as the univariational equations.  They are equivalent to the requirement that \eqref{VE1}, \eqref{VE2}, \eqref{VE3} hold for all curves $s\mapsto (p_\mu(t^i(s),q^\mu(t^i(s)),t^i(s))$.

The existence of solutions to the variational equations \eqref{usual_eqs1}, \eqref{usual_eqs2}, \eqref{usual_eqs3} implies that the functions $H_i$ Poisson commute.  To see this, we substitute \eqref{usual_eqs1} and \eqref{usual_eqs2} into \eqref{usual_eqs3}:
\begin{equation}
    0=\left(\frac{\partial H_j}{\partial p_\mu}\frac{\partial H_i}{\partial q^\mu}-\frac{\partial H_j}{\partial q^\mu}\frac{\partial H_i}{\partial p_\mu}\right)
    =\{H_j,H_i\}.
\end{equation}
Conversely, if the functions $H_i$ Poisson-commute then solutions $\Sigma$ to the variational problem exist, at least locally.  We prove this using the Frobenius integrability theorem.  The 2-form $dL$ defines a distribution $D\subset T(M\times\RR^n)$ such that $D_x=\{X\in T_x(M\times\RR^n)\::\:X\lrcorner dL=0\}$. The univariational principle is equivalent to the statement that $\Sigma$ is tangent to this distribution.  We will show that if $H_i$ Poisson-commute then the distribution is integrable and of rank $n$.  The Frobenius integrability theorem then ensures existence of solutions $\Sigma$ given by integral manifolds of the distribution.

The distribution is integrable because $dL$ is closed.  To see this, let $X,Y$ be two vector fields that take values in $D$.  This means that $X\lrcorner dL=Y\lrcorner dL=0$.  Then
\begin{equation}
    [X,Y]\lrcorner dL = \mathcal{L}_X (Y\lrcorner dL) - Y\lrcorner (\mathcal{L}_X dL) = 0-Y\lrcorner (d (X\lrcorner dL) + X_\lrcorner d^2L) = 0.
\end{equation}
So $[X,Y]$ takes values in $D$ and the distribution is integrable.
To see why the rank of $D$ is $n$, consider the one-forms
\begin{equation}
    \theta_\mu=-\frac{\partial}{\partial q^\mu}\lrcorner dL = dp_\mu+\frac{\partial H_i}{\partial q^\mu}dt^i,\quad
    \phi^\mu=\frac{\partial}{\partial p_\mu}\lrcorner dL = dq^\mu-\frac{\partial H_i}{\partial p_\mu}dt^i.
\end{equation}
These forms are linearly independent and generate a Pfaffian system.  Any tangent vector in $D$ is in the kernel of $\theta_\mu,\phi^\mu$ so the associated distribution contains $D$.  On the other hand,
\begin{equation}
    \theta_\mu\wedge\phi^\mu = dp_\mu\wedge dq^\mu -dH_i\wedge d t^i +\frac{\partial H_i}{\partial q^\mu}\frac{\partial H_j}{\partial p_\mu}dt^i\wedge dt^j = dL - \frac12\{H_i,H_j\}dt^i\wedge dt^j.
\end{equation}
Since $\{H_i,H_j\}=0$ this equation implies that any tangent vector in the kernel of $\theta_\mu,\phi^\mu$ belongs to $D$.  So the distribution $D$ is equivalent to the Pfaffian system $\{\theta_\mu,\phi^\mu\}$.  Therefore the rank of $D$ is $2m+n-2m=n$.  Since the distribution is integrable it admits an $n$-dimensional integrable submanifold $\Sigma$ which solves the univariational problem.

The univariational equations \eqref{usual_eqs1}, \eqref{usual_eqs2}, \eqref{usual_eqs3} were derived under the assumption that $\Sigma$ is the graph of a function $\RR^n\to M$.  Now we explain why this is not an assumption but a consequence of the univariational principle.  Any submanifold $\Sigma$ may be described parametrically by a map $s^i\mapsto(p_\mu(s^i),q^\mu(s^i),t^j(s^i))$ whose $n\times(2m+n)$ Jacobian matrix $(\partial p_\mu/\partial s^j,\partial q^\mu/\partial s^j,\partial t^i/\partial s^j)$ has full rank.  If $\Sigma$ solves the univariational principle then this map satisfies analogues of \eqref{usual_eqs1}, \eqref{usual_eqs2}, \eqref{usual_eqs3}:
\begin{align}
    \frac{\partial q^\mu}{\partial s^j}&=\frac{\partial t^i}{\partial s^j}\frac{\partial H_i}{\partial p_\mu}\,,\label{SE1}\\
    \frac{\partial p_\mu}{\partial s^j}&=-\frac{\partial t^i}{\partial s^j}\frac{\partial H_i}{\partial q^\mu}\,,\label{SE2}\\
    0&= \frac{\partial  q^\mu}{\partial s^j}\frac{\partial H_i}{\partial q^\mu}+ \frac{\partial p_\mu}{\partial s^j}\frac{\partial H_i}{\partial p_\mu}\,.\label{SE3}
\end{align}
These conditions imply that the $n\times n$ matrix $\partial t^i/\partial s^j$ is invertible.  To see why, suppose to the contrary: then there exists a nonzero vector $v^j$ such that $v^j\partial t^i/\partial s^j=0$.  Then equations \eqref{SE1} and \eqref{SE2} imply that $v^j\partial p_\mu/\partial s^j=0$ and $v^j\partial q^\mu/\partial s^j=0$.  This contradicts the assertion that the matrix $(\partial p_\mu/\partial s^j,\partial q^\mu/\partial s^j,\partial t^i/\partial s^j)$ has full rank, so $\partial t^i/\partial s^j$ must be invertible.  The inverse function theorem says that, since this matrix is invertible, we can find a local inverse $t^i\mapsto s^j(t^i)$ of the function $s^j\mapsto t^i(s^j)$.  This leads to a new parametrisation $t^i\mapsto (p_\mu(s^j(t^i)),q^\mu(s^j(t^i)),t^i)$.  So $\Sigma$ can always be parametrised as the graph of a function.

The action \eqref{geometric action} is written in local coordinates, so does not appear to be globally well-defined on $M$.  We end this section by explaining how to formulate the univariational principle globally on a symplectic manifold $(M,\omega)$.  For this, we need to choose two $m$-dimensional Lagrangian submanifolds $N_0,N_1\subset M$ and two vectors $t_0,t_1\in\RR^n$ which will determine boundary conditions.  We also choose a reference curve $\gamma_0:[0,1]\to M\times\RR^n$ such that $\gamma_0(0)\in N_0\times\{t_0\}$ and $\gamma_0(1)\in N_1\times\{t_1\}$.  Finally, we define
\begin{equation}
    \mathcal{S}[\gamma]=\int_\Delta (\omega-dH_i\wedge dt^i).
\end{equation}
Here $\Delta\subset M\times\RR^n$ is a surface whose boundary consists of four pieces: $\gamma_0$, $\gamma$, and two pieces contained in $N_0\times\{t_0\}$ and $N_1\times\{t_1\}$.  More precisely, $\Delta$ is the image of a map $\delta:[0,1]\times [0,1]\to\RR^n$ such that $\delta(0,s)=\gamma_0(s)$,  $\delta(1,s)=\gamma_1(s)$, $\delta(r,0)\in N_0\times\{t_0\}$ and $\delta(r,1)\in N_1\times\{t_1\}$.  This map $\delta$ is a homotopy connecting $\gamma_0$ to $\gamma$ that is based at $N_\alpha\times\{t_\alpha\}$.

Now consider a variation of $\gamma$.  This is described by a section $V$ of $T(M\times\RR^n)$ defined over $\Delta$.  This must vanish along $\gamma_0\subset\partial\Delta$ because $\gamma_0$ is fixed.  Along the pieces of $\partial\Delta$ contained in $N_\alpha\times\{t_\alpha\}$, $V$ must be tangent to $N_\alpha\times \{t_\alpha\}$ in order to preserve the boundary conditions of $\Delta$.  Since $\omega-dH_i\wedge dt^i$ is closed, the variation of $\mathcal{S}$ is given by
\begin{equation}
    \delta \mathcal{S} = \int_\Delta \mathcal{L}_V(\omega-dH_i\wedge dt^i)=\int_\Sigma d\iota_V(\omega-dH_i\wedge dt^i)=\int_{\partial\Delta}\iota_V(\omega-dH_i\wedge dt^i).
\end{equation}
This reduces the variation to a boundary integral.  We claim that the integral over three of the four boundary components is zero.  It is clearly zero on $\gamma_0$ because $V$ vanishes there.  Since $\omega-dH_i\wedge dt^i$ vanishes on $N_\alpha\times\{t_\alpha\}$, the integral is zero on these two parts of the boundary.  So the variation is given by an integral over the remaining piece $\gamma$ of the boundary:
\begin{equation}
    \delta \mathcal{S} = \int\gamma^\ast V\lrcorner(\omega-dH_i\wedge dt^i)= -\int_0^1 V\lrcorner \gamma'\lrcorner (\omega-dH_i\wedge dt^i) ds.
\end{equation}
The variational equation is therefore
\begin{equation} \gamma'\lrcorner (\omega-dH_i\wedge dt^i).\label{VE global}\end{equation}
This agrees with \eqref{VE} in local coordinates where $\omega=dp_\mu\wedge dq^\mu$.  The univariational principle then seeks an $n$-dimensional submanifold $\Sigma\subset M\times\RR^n$ such that \eqref{VE global} is satisfied by all curves $\gamma:[0,1]\to\Sigma$.

\section{Proof of equivalence}\label{sec:equivalence}


\subsection{Legendre transform} \label{Legendre}
Having presented the two variational problems, we now show that they are equivalent via a Legendre transform. The idea is of course as old as the Lagrangian/Hamiltonian formalism itself but we stress that the essential novelty (and complication) here is that we deal with Lagrangian one-forms, not the usual Lagrangian volume forms. The strategy of the proof is illustrated in the following diagram. 
\begin{center}
\begin{tikzcd}
& \mathscr{S}[q,v,\lambda,\Gamma] \arrow[dl, "(2)"'] \arrow[dr, "(1)", leftrightarrow] & \\
\mathcal{S}[\gamma] \arrow[rr, "(3)"'] & & S[q,\Gamma]
\end{tikzcd}    
\end{center}
Step (1) consists  establishing the equivalence between the position-space action $S[q,\Gamma]$ in \eqref{action traditional} and an extended action $\mathscr{S}[q,v,\lambda,\Gamma]$ introduced in \eqref{extended action} below. Step (2), we explain how to get our phase-space action of interest, $\mathcal{S}[\gamma]$ in 
\eqref{geometric action}, from $\mathscr{S}[q,v,\lambda,\Gamma]$. Finally, step (3) shows how to obtain $S[q,\Gamma]$ from the phase-space action $\mathcal{S}[\gamma]$.

Let us start with the traditional formulation of section \ref{traditional_formulation}.  We wish to convert the position-space action \eqref{action traditional} into the phase-space action \eqref{geometric action} using the Legendre transform.  To do so, it is convenient to introduce new variables $v^\mu_j$, which are not necessarily equal to the derivatives $\partial q^\mu/\partial t^j$, and $\lambda^j_{k\mu}$, which we will use as Lagrange multipliers in step $(1)$. We thus consider the following extended action
\begin{equation}\label{extended action}
    \mathscr{S}[q,v,\lambda,\Gamma]=\int_0^1 \left(L_k[q^\mu,v^\mu_j] + \lambda^j_{k\mu} \left(\frac{\partial q^\mu}{\partial t^j}-v^\mu_j\right)\right)\frac{dt^k}{ds}ds\,,
\end{equation}
where $q,v,\lambda$ are functions of $t^k$ and $\Gamma$ is a path in $\RR^n$ parametrised as $t^k(s)$, $s\in[0,1]$.

\textbf{Step (1)}: We ask that for all curves $\Gamma$, $\mathscr{S}$ is critical with respect to variations in $q,\lambda,\Gamma$.  Varying $\lambda^j_{k\mu}$ results in the constraint $v^\mu_j=\partial q^\mu/\partial t^j$ along the curve $\Gamma$.  If this holds for all curves $\Gamma$ then $v^\mu_j=\partial q^\mu/\partial t^j$ holds on all of $\RR^n$. Substituting back in $\mathscr{S}$ gives $S[q,\Gamma]$ and the rest of the bivariational principle is applied as in Section \ref{traditional_formulation}. Conversely, given the position-space action $S[q,\Gamma]$, it can always be trivially rewritten as the extended action $\mathscr{S}[q,v,\lambda,\Gamma]$ by treating $\lambda^j_{k\mu}$ as Lagrange multipliers. Therefore the variational problem for \eqref{extended action} is equivalent to that for \eqref{action traditional}.

\textbf{Step (2)}: We ask that for all curves $\Gamma$, $\mathscr{S}$ is critical with respect to variations in $q,v,\Gamma$. Similarly to the discussion in Section \ref{traditional_formulation}, considering variations $q^\mu\mapsto q^\mu+\delta q^\mu$ in \eqref{extended action}, we assume first that $\delta q^\mu=0$ along the curve $\Gamma$. The resulting variational equation is
\begin{equation}
    \lambda^j_{k\mu} = p_\mu \delta^j_k
\end{equation}
for some function $p_\mu$ of $t^j$, similar to \eqref{EL1}.  Next, varying $v$ results in the equation
\begin{equation}
\frac{\partial L_k}{\partial v^\mu_j}=\lambda^j_{k\mu}.
\end{equation}
The two equations together give
\begin{equation}\label{v variation}
    p_\mu \delta^j_k=\frac{\partial L_k}{\partial v^\mu_j}.
\end{equation}
We assume that this equation can be solved to write $v^\mu_j$ as a function of $p_\mu$ and $q^\mu$.  In general, \eqref{v variation} is an overdetermined equation for $v$, so asking that it can be solved imposes constraints on the Lagrangian coefficients $L_k$.  Inserting this solution back into \eqref{extended action} results in
\begin{equation}
\mathscr{S}[q,v,\lambda,\Gamma]\Big|_{    \lambda^j_{k\mu} = p_\mu \delta^j_k,v^\mu_j=v^\mu_j(p_\nu,q^\nu)}=\int_0^1 \left(p_\mu\frac{\partial q^\mu}{\partial t^j}-H_j(p_\nu,q^\nu)\right)\frac{dt^j}{ds}\,ds,
\end{equation}
in which
\begin{equation}
    H_j(p_\nu,q^\nu)=p_\mu v^\mu_j(p_\nu,q^\nu)-L_j(q^\nu,v^\nu_j(p_\lambda,q^\lambda)).
\end{equation}
This is an action of the form \eqref{geometric action}, in which the surface $\Sigma\subset M\times\RR^n$ is written as the graph of a function $(p_\mu(t^j),q^\mu(t^j))$.

\textbf{Step (3)}: Now we consider the reverse process, starting with the phase-space formulation of section \ref{symplectic_formulation} and aiming to recover the traditional formulation.  The degrees of freedom are an $n$-dimensional hypersurface $\Sigma\subset M\times\RR^n$.  We will assume that this is parametrised as a graph of a function $(t^1,\ldots,t^n)\mapsto (p_\mu(t^i),q^\mu(t^i))$.  No generality is lost, because (as explained above) any solution of the variational problem can be parametrised in this way.  Then the variational equations take the form \eqref{usual_eqs1}, \eqref{usual_eqs2}, \eqref{usual_eqs3}.

In classical mechanics, the inverse Legendre transform is performed by solving $\dot{q}^\mu=\partial H/\partial p_\mu$ to obtain $p_\mu$ as a function of $q^\nu$ and $\dot{q}^\nu$.  These are $m$ equations for $m$ unknowns, so it is reasonable to expect to find a solution.  The analogous equation \eqref{usual_eqs1} is $mn$ equations, so is overdetermined if regarded as an equation for $p_\mu$.  To circumvent this difficulty, we select one particular time direction.  Thus we fix a non-zero vector $\alpha\in\RR^n$ and seek to solve
\begin{equation}\label{alpha eq}
    \alpha^i q^\mu_i=\alpha^i\frac{\partial H_i}{\partial p_\mu},
\end{equation}
in which $q^\mu_i=\partial q^\mu/\partial t^i$.  We suppose that $\alpha$ can be chosen so that $\alpha^iH_i$ is a convex function of $p_\mu$.  Then the equation admits a unique solution $p_\mu=p_\mu(q^\nu,q^\nu_i)$.

To obtain the Lagrangian coefficients $L_i(q^\mu,q^\mu_i)$, we pull the one-form $L$ back to $\RR^n$ using the map $t^i\mapsto (p_\mu(t^i),q_\mu(t^i),t^i)$ and then substitute the solution $p_\mu=p_\mu(q^\nu,q^\nu_i)$ of \eqref{alpha eq}.  The pull-back of $L$ is
\begin{equation}
    p_\mu q^\mu_i\,dt^i - H_i(p_\nu,q^\nu)\,dt^i.
\end{equation}
So we obtain
\begin{equation}\label{Li}
    L_i = p_\mu(q^\nu,q^\nu_j)q^\mu_i - H_i(p_\nu(q^\sigma,q^\sigma_j),q^\nu).
\end{equation}
Thus, by solving one of the univariational equations (namely \eqref{alpha eq}) we have reduced the phase-space action \eqref{geometric action} to the traditional action \eqref{action traditional}.

Steps (1), (2), (3) put together give us the desired equivalence. There are two points to comment upon. 
First, we have described a Legendre transform that converts the traditional variational problem \eqref{action traditional} to the symplectic problem \eqref{geometric action}, and an inverse transform that goes the other way.  We must explain in what sense the ``inverse'' transform is the inverse of the Legendre transform.  Second, the inverse transform involves a choice of vector $\alpha$. We must show that this does not play any role in the set of variational equations or, equivalently, in the solution space.

Since the inverse transform involves a choice of $\alpha$, there are many position-space actions associated with a given phase space action.  So it is possible that applying the Legendre transform followed by the inverse transform to a given position-space action produces a different (but equivalent) position-space action.  Therefore the inverse transform is not a left inverse, but it is a right inverse, as we now explain.

Let us start with the phase-space action \eqref{geometric action} and apply the inverse Legendre transform followed by the Legendre transform.  We aim to show that we end up with the same action that we started with.  Applying the inverse Legendre transform results in a Lagrangian one-form \eqref{Li}, in which $p_\mu(q^\nu,q^\nu_j)$ is obtained by solving \eqref{alpha eq}.  To apply the Legendre transform to this, we introduce variables $v_i^\mu$ and consider the action \eqref{extended action} with
\begin{equation}\label{inverse L}
    L_k=p_\mu(q^\nu,v^\nu_i)v^\mu_k - H_k(p_\mu(q^\nu,v^\nu_i),q^\mu).
\end{equation}
As before, we vary $q$ and $v$ to obtain an equation similar to \eqref{v variation}:
\begin{equation}\label{inverse variations}
    \lambda^j_{k\mu}=\tilde{p}_\mu\delta^j_k=\frac{\partial L_k}{\partial v^\mu_j},
\end{equation}
in which the tilde distinguishes the new variable $\tilde{p}_\mu$ from the function $p_\mu(q^\nu,v^\nu_j)$.  We solve this to obtain $v^\mu_j$ as a function of $\tilde{p}_\mu$ and $q^\mu$.  To explicitly evaluate the right hand side we need to calculate $\partial p_\nu/\partial v^\mu_j$.  From equation \eqref{alpha eq} we obtain
\begin{equation}\label{inverse alpha equation}
    \alpha^i v^\mu_i=\alpha^i\frac{\partial H_i}{\partial p_\mu}\implies \alpha^j\delta^\mu_\nu = \alpha^i\frac{\partial^2 H_i}{\partial p_\mu\partial p_\lambda}\frac{\partial p_\lambda}{\partial v^\mu_j}.
\end{equation}
To solve this, let $g^{\mu\nu}=\alpha^i\partial^2H_i/\partial p_\mu\partial p_\nu$.  Assuming that $\alpha^i H_i$ is a convex function of $p$, we can invert $g$ to obtain $g_{\mu\nu}$ satisfying $g_{\mu\lambda}g^{\lambda\nu}=\delta_\mu^\nu$.  Then the equation is solved by
\begin{equation}
\label{eq_for_p}
    \frac{\partial p_\nu}{\partial v^\mu_j}=\alpha^jg_{\mu\nu}.
\end{equation}
We now evaluate the right hand side of \eqref{inverse variations} using \eqref{inverse L} and \eqref{eq_for_p}:
\begin{equation}
    \tilde{p}_\mu\delta^j_k=\frac{\partial L_k}{\partial v^\mu_j}=p_\mu\delta^j_k+\left(v^\nu_k-\frac{\partial H_k}{\partial p_\nu}\right)\frac{\partial p_\nu}{\partial v^\mu_j}
    =p_\mu\delta^j_k+\left(v^\nu_k-\frac{\partial H_k}{\partial p_\nu}\right)\alpha^j g_{\mu\nu}.
\end{equation}
Taking the trace of this equation and using \eqref{inverse alpha equation} shows that
\begin{equation}\label{inverse p}
    n\tilde{p}_\mu = np_\mu + \left(v^\nu_k-\frac{\partial H_k}{\partial p_\nu}\right)\alpha^k g_{\mu\nu}=np_\mu.
\end{equation}
Substituting back into the original equation then shows that $v^\nu_k$ is given in terms of $\tilde{p}_\mu,q^\mu$ by
\begin{equation}
    v^\nu_k=\frac{\partial H_k}{\partial p_\nu}\big(\tilde{p}_\mu,q^\mu\big).
\end{equation}
Substituting  \eqref{inverse L}, \eqref{inverse variations} and \eqref{inverse p} into \eqref{extended action} gives
\begin{equation}
    \mathscr{S}=\int_0^1 \left(p_\mu v^\mu_k-H_k(\tilde{p}_\mu,q^\mu)+\tilde{p}_\mu \left(\frac{\partial q^\mu}{\partial t^k}-v^\mu_k\right)\right)\frac{d t^k}{ds}\,ds = \int_0^1 \left(\tilde{p}_\mu \frac{\partial q^\mu}{\partial t^k}-H_k(\tilde{p}_\mu,q^\mu)\right)\frac{d t^k}{ds}\,ds.
\end{equation}
We thus recover the original action \eqref{geometric action} from the Legendre transform.

We now turn to the (absence of a) role of the choice of $\alpha$. As mentioned above, the solution $p_\mu$ depends on the choice of $\alpha$ and we make this explicit by writing $p_\mu=p_\mu(q^\nu,q^\nu_i,\alpha^i)$. On this solution, \eqref{alpha eq} becomes an identity and we can differentiate it with respect to $\alpha^k$ to obtain
\begin{equation}
\label{alpha_indep}
    q^\mu_k-\frac{\partial H_k}{\partial p_\mu}=\alpha^i\frac{\partial^2 H_i}{\partial p_\nu\partial p_\mu}\frac{\partial p_\nu}{\partial \alpha^k}.
\end{equation}
The left hand side is one of the equations of motion so it is zero ``on shell''.  The right hand side can be rewritten using the matrix $g^{\mu\nu}=\alpha^i\partial^2H_i/\partial p_\mu\partial p_\nu$, as in \eqref{inverse alpha equation}.  Assuming once again that this is invertible, the equation implies that $\frac{\partial p_\nu}{\partial \alpha^k}=0$.  Thus $p_\mu$ is independent of $\alpha$ on solutions of the equations of motion.  So the solutions of the equations of motion do not depend on $\alpha$. This will become clear in the examples below.  

\subsection{Example: the harmonic oscillator}
Let us illustrate all of this with a simple example.  Consider the two Hamiltonians
\begin{align}
    H_1 &= \frac12\left[\delta^{\mu\nu}p_\mu p_\nu + \delta_{\mu\nu}q^\mu q^\nu\right] \\
    H_2 &= \varepsilon^\mu_\nu p_\mu q^\nu,
\end{align}
in which $\mu,\nu$ run from 1 to 2, $\delta_{\mu\nu}$ is the Kronecker delta and $\varepsilon^\mu_\nu$ is totally antisymmetric with $\varepsilon^1_2=1$.  The first is the Hamiltonian of the two-dimensional harmonic oscillator and the second is the angular momentum associated with its rotational symmetry.  They generate the flows:
\begin{align}
p_{\mu 1} &= -\delta_{\mu\nu}q^\nu & p_{\mu 2} &= -\varepsilon^\nu_\mu p_\nu \\
q^\mu_1 &= \delta^{\mu\nu}p_\nu & q^\mu_2 &= \varepsilon^\mu_\nu q^\nu.
\end{align}
These are univariational equations for the phase-space action \eqref{geometric action}:
\begin{equation}\label{harmonic oscillator action}
    \mathcal{S}=\int_0^1 \left(p_\mu\frac{\partial q^\mu}{\partial t^i}-H_i\right)\frac{dt^i}{ds}ds.
\end{equation}
We wish to convert this to an action \eqref{action traditional} involving only position coordinates and no momentum coordinates.  We do so using the inverse Legendre transform.  Let us choose a non-zero vector $(\alpha^1,\alpha^2)\in\RR^2$.  Then equation \eqref{alpha eq} takes the form
\begin{equation}
    \alpha^1q_1^\mu+\alpha^2 q_2^\mu=\alpha^1\delta^{\mu\nu}p_\nu + \alpha^2\varepsilon^\mu_\nu q^\nu.
\end{equation}
Assuming that $\alpha^1\neq0$ and setting $\beta=\alpha^2/\alpha^1$, this is solved by
\begin{equation}
    p_\mu = \delta_{\mu\nu}q_1^\nu + \beta\delta_{\mu\nu}\left(q_2^\nu-\varepsilon^\nu_\rho q^\rho\right).
\end{equation}
Note that, although $p_\mu$ depends on $\beta$, it is independent of $\beta$ on solutions of the univariational equations, as explained above.  Substituting into \eqref{harmonic oscillator action} gives the position-space Lagrangian one-form 
\begin{align}
    L_1 &= \frac{1}{2}\left[\delta_{\mu\nu}q_1^\mu q_1^\nu - \beta^2\delta_{\mu\nu}(q_2^\mu-\varepsilon^\mu_\rho q_2^\rho)(q_2^\nu-\varepsilon^\nu_\sigma q_2^\sigma)-\delta_{\mu\nu}q^\mu q^\nu\right]\\
    L_2 &= \delta_{\mu\nu}\left[q_1^\mu+\beta(q_2^\mu-\varepsilon^\mu_\rho q^\rho)\right]\left[q_2^\nu-\varepsilon^\nu_\sigma q^\sigma\right].
\end{align}
Although this depends on $\beta=\alpha^2/\alpha^1$, the solutions of the multi-time Euler--Lagrange equations are independent of $\beta$.

\subsection{Example: periodic Toda chain}
Consider the two Hamiltonians
\begin{align}
    H_1 &= \sum_\mu\frac12(p_\mu)^2 + \exp(q^\mu-q^{\mu-1})\\
    H_2 &= \sum_\mu\frac13(p_\mu)^3+(p_\mu+p_{\mu-1})\exp(q^\mu-q^{\mu-1}).
\end{align}
These depend on $2m$ variables $q^1,\ldots,q^m,p_1,\ldots,p_m$, with indices understood modulo $m$.  It is straightforward to check that
\begin{equation}
    \frac{\partial H_1}{\partial p_\mu}\frac{\partial H_2}{\partial q^\mu}=\sum_\mu(p_\mu)^2\big[\exp(q^\mu-q^{\mu-1})-\exp(q^{\mu+1}-q^\mu)\big] = \frac{\partial H_2}{\partial p_\mu}\frac{\partial H_1}{\partial q^\mu}
\end{equation}
and so the two Hamiltonians Poisson commute.  The Hamiltonian $H_1$ describes the Toda lattice and $H_2$ represents a conserved quantity.  The phase-space action \eqref{geometric action} for this system is
\begin{equation}
    S=\int_0^1 \left(p_\mu\frac{\partial q^\mu}{\partial t^i}-H_i\right)\frac{dt^i}{ds}ds.
\end{equation}
We wish to convert this to an action \eqref{action traditional} involving only position coordinates and no momentum coordinates.  We do so using the inverse Legendre transform.  Let us choose a non-zero vector $(\alpha^1,\alpha^2)\in\RR^2$, and assuming that $\alpha^1\neq0$, set $\beta=\alpha^2/\alpha^1$. Then equation \eqref{alpha eq} takes the form
\begin{equation}
    q_1^\mu+\beta q_2^\mu=p_\mu+\beta(p_\mu)^2+\beta\exp(q^{\mu+1}-q^\mu)+\beta\exp(q^{\mu}-q^{\mu-1}).
\end{equation}
This is a set of $m$ quadratic equations for $p_\mu$.  They are solved by
\begin{equation}\label{toda momentum}
    p_\mu = \frac{2q_1^\mu+2\beta( q_2^\mu-\exp(q^{\mu+1}-q^\mu)-\exp(q^{\mu}-q^{\mu-1}))}{1\pm\sqrt{1+4\beta q_1^\mu+4\beta^2(q_2^\mu-\exp(q^{\mu+1}-q^\mu)-\exp(q^{\mu}-q^{\mu-1}))}}
\end{equation}
We obtain a Lagrangian multiform by substituting these expressions into:
\begin{align}
    L_1 &= \sum_\mu p_\mu q^1_\mu - \frac12(p_\mu)^2-\exp(q^\mu-q^{\mu-1}) \\
    L_2 &= \sum_\mu p_\mu q_2^\mu - \frac13(p_\mu)^3-(p_\mu+p_{\mu-1})\exp(q^\mu-q^{\mu-1})
\end{align}
Thus we have a family of Lagrangian multiforms for this system, parametrised by $\beta\in\mathbb{R}$ together with $m$ sign choices in \eqref{toda momentum}.  If we choose all plus signs in \eqref{toda momentum} and set $\beta=0$ the Lagrangian multiform is
\begin{align}
    L_1 &= \sum_\mu \frac12 (q_1^\mu)^2-\exp(q^\mu-q^{\mu-1}) \\
    L_2 &= \sum_\mu q_1^\mu q_2^\mu - \frac13(q_1^\mu)^3-(q_1^\mu+q_1^{\mu-1})\exp(q^\mu-q^{\mu-1}).
\end{align}
This recovers the Lagrangian coefficients presented in \cite{PS} from our more general perspective.

\subsection{Linear dependence on velocities}

The Legendre transform introduced in section \ref{Legendre} relates an action for a traditional Lagrangian multiform to a phase-space action by solving equation \eqref{v variation} for the velocities $v^\mu_j$.  But if the Lagrangian coefficients $L_k$ depend linearly on velocities the right hand side of \eqref{v variation} is independent of velocity, so this equation cannot be solved.  The question arises as to whether the equivalence still holds in the case where the Lagrangian coefficients are linear in velocities.  In this section we address this question.

If a Lagrangian multiform is linear in velocities then the variational equation \eqref{equal momenta} implies that it takes the form
\begin{equation}
    L_k = p_\mu(q^\nu)q^\mu_k - V_k(q^\nu)
\end{equation}
for some functions $p_\mu$ and $V_k$ of $q^\nu$.  Then the action \eqref{action traditional} takes the form $S=\int\gamma^\ast L$, in which
\begin{equation}\label{linear multiform}
    L = p_\mu(q^\nu)d q^\mu - V_k(q^\nu)dt^k
\end{equation}
and $\gamma$ is a function of the form $s\mapsto (q^\mu(t^i(s)),t^k(s))$.  This similar in form to the action \eqref{geometric action}, but an important difference is that here $p_\mu$ are functions of the coordinates $q^\nu$, whereas in \eqref{geometric action} $p_\mu$ and $q^\mu$ are independent coordinates.
To directly compare this with the action \eqref{geometric action} we consider the 2-form
\begin{equation}
\label{Omega}
    \Omega = dp_\mu\wedge dq^\mu = \frac{\partial p_\mu}{\partial q^\nu}dq^\nu\wedge dq^\mu.
\end{equation}
The rank of this 2-form at a point $q$ is the rank of the linear map $X=X^\mu\partial_\mu\mapsto X\lrcorner\Omega = X^\mu (\partial_\mu p_\nu-\partial_\nu p_\mu)dq^\mu$.  The form $\Omega$ is called nondegenerate if this map has full rank at every point $q$.

Suppose that $\Omega$ is nondegenerate.  Then it is by definition a symplectic form and the dimension $m$ is even.  By the Darboux theorem, we can choose coordinates $P_\mu,Q^\mu$ in which $\Omega=\sum_{\mu=1}^{m/2}dP_\mu \wedge dQ^\mu$.  In these coordinates, the action is precisely in the form of a phase-space action \eqref{geometric action}.

This exact situation arose in \cite{CDS,CSV}, which produced classes of Lagrangian one-forms which are linear in the velocities.  In all of those cases, the 2-form $\Omega$ \eqref{Omega} is nondegenerate since it corresponds to the (pull-back of the) Kostant-Kirillov symplectic form on an appropriate coadjoint orbit. It was shown in \cite{CDS,CSV} how suitably parametrising the coadjoint orbit yields Darboux canonical coordinates.

The case where $\Omega$ is degenerate is more involved.  The case of a single time coordinate was discussed in \cite{FJ}, which showed that the variational problem can be reduced to one for which $\Omega$ is nondegenerate. The corresponding analysis of degeneracy in the present context of multiple time coordinates is much more complicated and beyond the scope of this article.

\section{Phase-space Lagrangian one-forms on a Lie group}\label{sec:form_group}

In this section we present a generalisation of the univariational principle, in which multi-time $\RR^n$ is replaced by a Lie group. This idea was first appeared in \cite{CNSV} with the view to incorporate super integrable systems into the theory of Lagrangian multiforms. Doing so, we establish for the first time that the equations describing (infinitesimal) Hamiltonian group actions of a (connected) Lie group on a symplectic manifold are actually variational. They appear as the Euler-Lagrange equations of our univariational principle applied to a natural generalisation of the Lagrangian one-form \eqref{basis_one_form}. We also show that the same results, reexpressed in local group coordinates, can be interpreted as describing a collection of compatible {\it non autonomous} Hamiltonian equations. In particular, this accommodates integrable systems with explicitly time-dependent constants of motion. 

\subsection{A simple motivating example}\label{simple}

To understand and motivate our construction, let us consider the 2D harmonic oscillator.
Here the symplectic manifold is simply $\RR^4$ with canonical coordinates $(q_1,q_2,p_1,p_2)$ and symplectic form $\omega=dp_1\wedge dq_1+dp_2\wedge dq_2$. The 2D (isotropic) harmonic oscillator Hamiltonian is given by
$$H=H_1+H_2\,,~~H_j=\frac{1}{2}(p_j^2+q_j^2)\,,~~j=1,2\,.$$
Consider the following 3 functions on the phase space
\begin{equation}
    J_1=H_1-H_2\,,~~J_2=p_1q_2-p_2q_1\,,~~J_3=p_1p_2+q_1q_2\,.
\end{equation}
The following facts are well-known. The system is integrable since, for instance, $\{H,J_2\}=0$: there are two functionally independent first integrals. It is superintegrable since, additionally, we also have $\{H,J_1\}=0$ or $\{H,J_3\}=0$. The functions $J_1,J_2,J_3$ provide a realisation of the ${\rm su}(2)$ Lie algebra
\begin{equation}
\label{rels_J}
    \{J_i,J_j\}=2 \varepsilon_{ijk}J_k\,,
\end{equation}
which is therefore a symmetry of the 2D harmonic oscillator. Note the relation $J_1^2+J_2^2+J_3^2=H^2$.

Using the results of the previous section, we can easily produce a Lagrangian multiform for the integrable system $H,J_2$ say. 
It suffices to define 
\begin{equation}
    L=p_\mu dq^\mu-H dt^1-J_2dt^2.
\end{equation}
Then the system of Euler-Lagrange equations is as in \eqref{usual_eqs1}-\eqref{usual_eqs2}, with $t^1$ associated with $H$ and $t^2$ with $J_2$. The compatibility of the system, encoded in \eqref{usual_eqs3}, holds
since $\{H,J_2\}=0$.

We would like to go further and produce a Lagrangian multiform and univariational principle not only for the integrable system $H,J_2$ but for $H$ {\it together with} its entire symmetry algebra $J_1,J_2,J_3$. Note in particular that, if feasible, this would produce a Lagrangian multiform for the superintegrable system $H,J_2,J_3$ say. 

The main difficulty to overcome is the non abelian algebra generated by $J_1,J_2,J_3$. It can no longer be expected that we can consider compatible systems associated to $J_1$ and $J_2$ for instance. Similarly, the closure relation, if still valid in some appropriate form, can no longer be expected to be the variational analog of the Poisson involutivity of the functions $J_1,J_2,J_3$ since the latter are not in involution. Instead, it should produce the appropriate relations \eqref{rels_J}.

\subsection{A general construction}
\label{sec:group general}

The basic ideas to produce Lagrangian one-forms for non-commuting conserved quantities were laid out in \cite{CNSV}. Here, we formulate them systematically in the geometric framework of Section \ref{symplectic_formulation}. In doing so we elucidate the underlying geometric feature that ensures that the closure relation in the non abelian setting yields the correct equations: it has to do with the Maurer-Cartan equation for (left) invariant one-forms on an appropriate Lie group $G$.

Let $M$ be a $2m$-dimensional symplectic manifold and $G$ a connected Lie group of dimension $n$ with Lie algebra $\g$.  We assume for convenience that the symplectic form on $M$ is exact and that it can be written $\omega=d\alpha$, with $\alpha=p_\mu dq^\mu$ in suitable coordinates. We showed in Section \ref{symplectic_formulation} how to proceed if this is not the case. Let $H:M\to \g^*$ be a smooth map and consider the Lagrangian one-form
\begin{equation}
\label{Lag_on_G}
    L=\alpha-(H,g^{-1}dg)\,,
\end{equation}
where $(~,~)$ is the pairing between $\g^*$ and $\g$. 
The left-invariant Maurer-Cartan one-form $g^{-1}dg$ can be written as $g^{-1}dg=E_i\otimes \theta^i$ where $E_i$ is a basis for $\g$ and $\theta^i$ the dual basis of left-invariant one-forms on $G$ \footnote{These are dual in the sense that $\theta^i(e)\in T^\ast_e G\cong\g^\ast$ satisfy $(\theta^i(e),E_j)=\delta^i_j$.}. Thus, we can also write
\begin{equation}\label{general_L}
    L=p_\mu dq^\mu-H_i\theta^i\,,~~H_i=(H,E_i)\,.
\end{equation}
For any parametrised curve $\gamma:[0,1]\to M\times G$ we define the associated action as
\begin{equation}
    S[\gamma]=\int_0^1\gamma^\ast L.
\end{equation}
By analogy with section \ref{symplectic_formulation}, we introduce a univariational principle, which demands existence of a submanifold $\Sigma\subset M\times G$ such that every curve $\gamma\subset \Sigma$ is a critical point of $S[\gamma]$ and $\dim\Sigma=\dim G$.

The Euler-Lagrange equations for $S$ again take the form $\gamma'\lrcorner dL = 0$; let us write these out more explicitly.  By direct calculation,
\begin{equation}
    dL=dp_\mu\wedge dq^\mu-\frac{\partial H_j}{\partial q^\mu}dq^\mu\wedge\theta^j-\frac{\partial H_j}{\partial p_\mu}dp_\mu\wedge\theta^j-H_j\,d\theta^j \,.
\end{equation}
The key point now is that the last term can be evaluated using the Maurer-Cartan equation 
\begin{equation}
    d\theta^i=-\frac{1}{2}c_{jk}^i\,\theta^j\wedge\theta^k\,,
\end{equation}
in which the structure constants $c_{jk}^i$ are defined by $[E_j,E_k]=c_{jk}^iE_i$.
With $\gamma'\lrcorner dp_\mu=\frac{dp_\mu}{ds}$, $\gamma'\lrcorner dq^\mu=\frac{dq^\mu}{ds}$ and $\gamma'\lrcorner \theta^i=Y^i$,
we obtain
\begin{align}
    \gamma'\lrcorner dL=\left(  -  \frac{dq^\mu}{ds}+\frac{\partial H_j}{\partial p_\mu}Y^j  \right)dp_\mu+\left(      \frac{dp_\mu}{ds}+\frac{\partial H_j}{\partial q^\mu}Y^j\right)dq^\mu-\left(\frac{d q^\mu}{ds}\frac{\partial H_i}{\partial q^\mu}+\frac{d p_\mu}{ds}\frac{\partial H_i}{\partial p_\mu}-c^\ell_{ji}H_\ell Y^j  \right)\theta^i
\end{align}
This gives the following generalisation of \eqref{VE1}-\eqref{VE3}
\begin{align}
    \frac{dq^\mu}{ds}&=\frac{\partial H_j}{\partial p_\mu}Y^j\,,\label{genEL1}\\
    \frac{dp_\mu}{ds}&=-\frac{\partial H_j}{\partial q^\mu}Y^j\,,\label{genEL2}\\
    0&= \frac{d q^\mu}{ds}\frac{\partial H_i}{\partial q^\mu}+\frac{d p_\mu}{ds}\frac{\partial H_i}{\partial p_\mu}-c^\ell_{ji}H_\ell Y^j \label{genEL3}\,.
\end{align}
By substituting \eqref{genEL1} and \eqref{genEL2} into \eqref{genEL3} we deduce from these Euler--Lagrange equations that
\begin{equation}\label{genEL4}
\left(\frac{\partial H_j}{\partial p_\mu}\frac{\partial H_i}{\partial q^\mu}-\frac{\partial H_j}{\partial q^\mu}\frac{\partial H_i}{\partial p_\mu}-c^\ell_{ji}H_\ell\right) Y^j=0\,.
\end{equation}

The univariational principle demands that equations \eqref{genEL1}--\eqref{genEL4} hold for all curves $\gamma$ in $\Sigma$.  By similar arguments to those presented in section \ref{symplectic_formulation}, this means that \eqref{genEL4} holds for all $Y^j$.  Thus
\begin{equation}\label{Hc}
    \{H_i,H_j\}=c_{i j}^kH_k\,,
\end{equation}
where we used the Poisson bracket 
$$\{F,G\}=\frac{\partial F}{\partial p_\mu}\frac{\partial G}{\partial q^\mu}-\frac{\partial G}{\partial p_\mu}\frac{\partial F}{\partial q^\mu}\,.$$
Equation \eqref{Hc} can be used to encode the Poisson brackets of the superintegrable system of Section \ref{simple} that we presented as motivation for the present general construction. Indeed, if we choose $G=\RR\times SU(2)$ then \eqref{Hc} describes the Poisson brackets of the conserved quantities $H,J_1,J_2,J_3$ for the 2D harmonic oscillator.

Equations \eqref{Hc}, which follow from the univariational principle, imply that $H:M\to\g^\ast$ is a moment map for a Hamiltonian action of $G$ on $M$.  Moment maps arise in the theory of symplectic quotients and it is interesting to compare our Lagrangian one-form with the symplectic quotient construction.  Suppose that there is a Hamiltonian action of $G$ on $M$ with moment map $\phi:M\to\g^\ast$.  Now consider the symplectic manifold $T^\ast G$.  This can be identified with $\mathfrak{g}^\ast\times G$ in such a way that the tautological one-form is written $(f,g^{-1}dg)$ for $f\in\mathfrak{g}^\ast$ and $g\in G$.  Let
\begin{equation}
\beta = p_\mu dq^\mu + (f,g^{-1}dg).
\end{equation}
Then $d\beta$ is a symplectic form on $M\times T^\ast G$. There is a natural action of $G$ on $M\times T^\ast G$ induced by the action on $M$ and the left action on $G$.  The moment map for this action is
\begin{equation}
\Phi(p_\mu,q^\mu,f,g)=\phi(p_\mu,q^\mu)+f.
\end{equation}
The symplectic quotient is the quotient of $\Phi^{-1}(0)$ by $G$.  Now $\Phi^{-1}(0)\subset M\times T^\ast G$ is identified with $M\times G$ by the natural projection $M\times T^\ast G\to M\times G$.  Under this identification, the one-form $\beta$ becomes
\begin{equation}
\beta = p_\mu dq^\mu - (\phi,g^{-1}dg).
\end{equation}
If we identify $\phi$ with $H$ this takes the form of the Lagrangian 1-form $L$ on $M\times G$.  Moreover, the surfaces $\Sigma$ in the univariational principle are precisely the orbits of the action of $G$ on $\Phi^{-1}(0)$. So the theory of symplectic quotients offers a natural interpretation of the structure of our Lagrangian one-form.  However, we stress that the starting points of the two constructions are different. Following the philosophy of variational principles, we choose the one-form $L$ and postulate the univariational principle as the starting point. At that stage, our map $H$ is not yet a moment map and the group $G$ does not act on $M$ by Hamiltonian action. The application of the univariational principle then produces ``equations of motion'' which tell us that there is a Hamiltonian action of $G$ on $M$ and that $H$ is the associated moment map. In contrast, the theory of symplectic quotients takes the existence of a Hamiltonian action and associated moment map as the starting point. In other words, we have demonstrated that a Hamiltonian group action on a symplectic manifold derives from our univariational principle. 

\subsection{Explicitly time-dependent commuting flows}

It is instructive to rewrite the variational equations \eqref{genEL1}, \eqref{genEL2}, \eqref{genEL3} in local coordinates on $G$, say $\tau^k$, $k=1,\dots,n$. Each left-invariant one-form can be written in the basis of coordinate one-form as
\begin{equation}
\label{matrix_theta}
	\theta^j=\theta^j_kd\tau^k\,,
\end{equation}
where the coefficients $\theta^j_k$ are (smooth) functions of $\tau^\ell$. Note that the matrix of coefficients $\theta^j_k$ is invertible, since this is a change of basis. Details on how to compute $\theta^j_k$ in terms of the structure constants of the Lie algebra and relative to a choice of coordinates are given for instance in \cite{MMS}. With \eqref{matrix_theta}, we have $Y^j=\theta^j_k\frac{d\tau^k}{ds}$ and \eqref{genEL1}-\eqref{genEL3} read
\begin{align}
	\frac{dq^\mu}{ds}&=\frac{\partial H_j}{\partial p_\mu}\theta^j_k\frac{d\tau^k}{ds}\,,\label{GH1}\\
	\frac{dp_\mu}{ds}&=-\frac{\partial H_j}{\partial q^\mu}\theta^j_k\frac{d\tau^k}{ds}\,,\label{GH2}\\
	0&= \frac{d q^\mu}{ds}\frac{\partial H_i}{\partial q^\mu}+\frac{d p_\mu}{ds}\frac{\partial H_i}{\partial p_\mu}-c^\ell_{ji}H_\ell \theta^j_k\frac{d\tau^k}{ds} \label{GH3}\,.
\end{align}
It is convenient to multiply the last equation by $\theta^i_m$, to introduce the functions
\begin{equation}
	\label{def_K}
	K_k=H_j\theta^j_k\,,~~k=1,\dots,n\,,
\end{equation}
and to use the following consequence of the Maurer-Cartan equation on the coefficients $\theta^j_k$
\begin{equation}
	\label{MC_eq_comp}
	\frac{\partial \theta^\ell_r}{\partial \tau^m}-\frac{\partial \theta^\ell_m}{\partial \tau^r}=c_{ji}^\ell\theta^j_k\theta^i_m\,,
\end{equation}
in order to rewrite \eqref{GH1}-\eqref{GH3} equivalently as
\begin{align}
	\frac{dq^\mu}{ds}&=\frac{\partial K_k}{\partial p_\mu}\frac{d\tau^k}{ds}\,,\label{GSE1K}\\
	\frac{dp_\mu}{ds}&=-\frac{\partial K_k}{\partial q^\mu}\frac{d\tau^k}{ds}\,,\label{GSE2K}\\
	0&= \left(\frac{\partial K_k}{\partial p_\mu}\frac{\partial K_m}{\partial q^\mu}-\frac{\partial K_k}{\partial q^\mu} \frac{\partial K_m}{\partial p_\mu}-\left(\frac{\partial K_k}{\partial \tau^m}-\frac{\partial K_m}{\partial \tau^k} \right)\right)\frac{d\tau^k}{ds}\label{GSE3K}\,.
\end{align}
Summarising, this system of equations arises as the Euler-Lagrange equations of our Lagrangian one-form which now reads
\begin{equation}
	\label{Lag_K}
	L=p_\mu dq^\mu-K_j d\tau^j\,.
\end{equation}

The univariational principle demands that \eqref{GSE1K}, \eqref{GSE2K}, \eqref{GSE3K} hold for all curves $\tau^j(s)$.  This leads to the following system of equations for functions $p_\mu(\tau^k)$, $q^\mu(\tau^k)$:
\begin{align}
	\frac{\partial q^\mu}{\partial \tau^k}&=\theta^j_k(\tau^i)\big\{H_j(p_\nu,q^\nu),q^\mu\big\}\label{non_auto1}\,,\\
	\frac{\partial p_\mu}{\partial \tau^k}&=\theta^j_k(\tau^i)\big\{H_j(p_\nu,q^\nu),p_\mu\big\}\label{non_auto2}\,,\\
    0&=\theta_k^i\theta_m^j\{H_i,H_j\}-\left(\frac{\partial\theta_k^i}{\partial \tau^m}-\frac{\partial\theta_m^i}{\partial \tau^k}\right)H^i\,,\label{non_auto3}
\end{align}
where we have recalled explicitly the coordinate dependence in the first two equations.
The latter equations describe a collection of flows in time directions $\tau^k$.  Interestingly, these flows are non-autonomous, because the time variables $\tau^k$ appear explicitly on the right hand side. The third equation is of course equivalent to \eqref{Hc} (recalling \eqref{MC_eq_comp}) but is more convenient in this form to check directly the consistency for these flows. To see this, let $f$ be a function of $p_\mu,q^\mu$ and suppose that $p_\mu(\tau^i),q^\mu(\tau^i)$ solve \eqref{non_auto1}, \eqref{non_auto2}.  Then
\begin{align}
	\frac{\partial}{\partial\tau^j}\frac{\partial}{\partial\tau^k}f&=\frac{\partial\theta^\ell_k}{\partial\tau^j}\{H_\ell,f\}+\theta^\ell_k\frac{\partial}{\partial\tau^j}\{H_\ell,f\}\nonumber\\
    &=\frac{\partial\theta^\ell_k}{\partial\tau^j}\Big\{H_\ell,f\Big\}+\theta^\ell_k\theta^m_j \{H_m,\{H_\ell,f\}\}\nonumber
\end{align}
Consistency requires that
\begin{equation}
    0=\frac{\partial}{\partial\tau^j}\frac{\partial}{\partial\tau^k}f-\frac{\partial}{\partial\tau^k}\frac{\partial}{\partial\tau^j}f
    =\left(\frac{\partial\theta^\ell_k}{\partial\tau^j}-\frac{\partial\theta^\ell_j}{\partial\tau^k}\right)\Big\{H_\ell,f\Big\}+\theta^\ell_k\theta^m_j \big(\{H_m,\{H_\ell,f\}\}-\{H_\ell,\{H_m,f\}\}\big).
\end{equation}
By the Jacobi identity for the Lie bracket, this is equivalent to \eqref{non_auto3}.

The non-autonomous equations \eqref{non_auto1}, \eqref{non_auto2} provide a framewoork that can accommodate systems with constants of motion with explicit time-dependence. We illustrate this with an example. Let $M=\RR^2$ with the canonical form $\omega=dp\wedge dq$, and the Hamiltonian (for some fixed $a>0$),
$$H_0=\frac{p^2}{2m}+\frac{a}{q^2}\,.$$
Being a one degree of freedom system, it is trivially integrable. 
With $J=\frac{pq}{2}$, direct calculation gives
$$\{H_0,J\}=H_0$$
so that $C=J-tH_0$ is conserved under the flow of $H_0$: 
\begin{equation}
\label{conservation_J}
    \frac{dC}{dt}=\{H_0,C\}+\frac{\partial C}{\partial t}=H_0-H_0=0\,.
\end{equation}

To cast this in our framework, consider $G$ the Lie group of $2\times 2$ upper triangular matrices with unit determinant. A basis of the Lie algebra is given by
$$\xi_1=\begin{pmatrix}
	0 & 1\\
	0 & 0
\end{pmatrix}\,,~~\xi_2=\frac{1}{2}\begin{pmatrix}
	-1 & 0\\
	0 & 1
\end{pmatrix}$$
with $[\xi_1,\xi_2]=\xi_1$. We parametrise an element $g$ of $G$ using coordinates $\tau_1=t$, $\tau_2=\tau$ as follows
\begin{equation}
    g=e^{\tau \xi_2}e^{t\xi_1}=\begin{pmatrix}
        e^{-\tau/2} & t e^{-\tau/2}\\
        0 & e^{\tau/2}
    \end{pmatrix}\,.
\end{equation}
Thus, we have 
$$g(t,\tau)^{-1}dg(t,\tau)=\xi_1\otimes dt+(\xi_2-t\xi_1)\otimes d\tau=\xi_1\otimes\theta^1+\xi_2\otimes\theta^2\,,$$
giving in particular the left-invariant one-forms as $\theta^1=dt-td\tau$, $\theta^2=d\tau$.
Let $\{\mu^1,\mu^2\}$ be the basis in $\g^*$ dual to $\{\xi_1,\xi_2\}$.  Using the non degenerate bilinear form $\langle \xi,\eta\rangle={\rm Tr}(\xi\eta)$ on $\g$, we have
$$\mu^1=\begin{pmatrix}
    0 & 0\\
    1 & 0
\end{pmatrix}\,,~~\mu^2=\begin{pmatrix}
    -1 & 0\\
    0 & 1
\end{pmatrix}\,.$$
With the map $H:M\to \g^*$ written as  
$$H=H_1\mu^1+H_2\mu^2\,;$$
the Lagrangian one-form \eqref{Lag_on_G}, equivalently \eqref{Lag_K}, reads
 \begin{equation}
 \label{time_dep_Lag}
     L=pdq-\left(H\,,\, g^{-1}dg \right)=pdq-H_1dt-(H_2-tH_1) d\tau=pdq-K_1dt-K_2 d\tau\,.
 \end{equation}
The connection with the above simple physical example is now immediate if we identify $H_1=H_0(=K_1)$, $H_2=J$ so that $(H_2-tH_1)=C(=K_2)$. The system of (non autonomous) compatible equations obtained from \eqref{non_auto1}-\eqref{non_auto2} with $\tau_1=t$ and $\tau_2=\tau$ reads
\begin{equation}
	\begin{cases}
		\frac{\partial q}{\partial t}=\frac{p}{m}\,,\\
		\frac{\partial p}{\partial t}=\frac{2a}{q^3}\,,
	\end{cases}\,,~~
	\begin{cases}
	\frac{\partial q}{\partial\tau}=\frac{q}{2}-t\frac{p}{m}\,,\\
	\frac{\partial p}{\partial\tau}=-\frac{p}{2}-t\frac{2a}{q^3}\,.
\end{cases}
\end{equation}
We recover by direct calculation that $\frac{\partial K_2}{\partial t}=0$ and now also that $\frac{\partial K_1}{\partial\tau}=-K_1$. Thus, for this simple example, the solutions $q(t,\tau)$, $p(t,\tau)$ can be described by the equations
$$\frac{p^2}{2m}+\frac{a}{q^2}=c_1e^{-\tau}\,,~~pq=2(c_2+c_1t)\,,~~c_1,c_2\in\RR\,.$$

\subsection{Lagrangian one-form from matrix representations of Lie algebras}

We can form a Lagrangian one-form associated to any matrix representation on $V$ of any finite-dimensional Lie algebra $\g$ as follows. Suppose we have a matrix representation 
$$[M_i,M_j]=c_{ij}^kM_k\,,~~M_i\in {\rm End(V)}\subseteq{\rm Mat}_m(\CC)\,,~~i=1,\dots,n\,,$$
and that $M$ is a symplectic manifold with canonical coordinates $p_\mu,q^\mu$, $\mu=1,\dots,m$,
$$\{q^\mu,q^\nu\}=0=\{p_\mu,p_\nu\}\,,~~\{p_\mu,q^\nu\}=\delta_\mu^\nu\,.$$
The following simple well-known observation
$$\{p_\mu A^\mu_\nu q^\nu,p_\sigma B^\sigma_\xi q^\xi\}=- p_\mu [A,B]^\mu_\nu q^\nu$$
for any matrices $A,B$ allows us to define 
\begin{equation}
	\label{realisation_H}
	H_i=-p_\mu (M_i)^\mu_\nu q^\nu 
\end{equation}
to obtain the following canonical realisation of the Lie algebra $\g$
$$\{H_i,H_j\}=c_{ij}^kH_k\,.$$
It remains to set $L$ as in \eqref{general_L} to obtain a desired Lagrangian one-form with the desired properties.  Additionally, note that 
$$\{ p_\mu A^\mu_\nu q^\nu ,p_\sigma \}=-p_\mu A^\mu_\sigma\,.$$
This makes it possible to realise canonically certain Lie algebras of semi-direct groups, such as the Poincar\'e group.
Its $10$-dimensional Lie algebra can be conveniently written in terms of the six generators of the Lorentz transformations $L_{\alpha\beta}=-L_{\beta\alpha}$\footnote{It is more convenient for this example to work with generators $L_{\alpha\beta}$ with two indices, modulo the antisymmetry relation, rather than with $L_i$ as suggested by our general discussion. We will change the notation accordingly for the matrix representation $M_{\alpha\beta}$ used for illustration.}, $\alpha,\beta=0,1,2,3$ and the four generators of translations $P_\mu$, $\mu=0,1,2,3$:
\begin{align}
    &[L_{\alpha\beta},L_{\gamma\xi}]=\eta_{\beta\gamma}L_{\alpha\xi}-\eta_{\alpha\gamma}L_{\beta\xi}+\eta_{\alpha\xi}L_{\beta\gamma}-\eta_{\beta\xi}L_{\alpha\gamma}\,,\\
    &[L_{\alpha\beta},P_\mu]=\eta_{\beta\mu}P_\alpha-\eta_{\alpha\mu}P_\beta\,,\\
    &[P_\mu,P_\nu]=0\,,
\end{align}
where $\eta={\rm diag}(-1,+1,+1,+1)$. For conciseness, let us restrict our attention to the Lorentz subalgebra and consider the following $4$-dimensional representation $M_{\alpha\beta}$ for the Lorentz generators $L_{\alpha\beta}$, with matrix elements
\begin{equation}
	\label{representation_L}
	(M_{\alpha\beta})^\mu_\nu=\delta_\alpha^\mu\eta_{\beta\nu}-\delta_\beta^\mu\eta_{\alpha\nu}\,.
\end{equation}
The realisation \eqref{realisation_H} thus gives the following components of the map $H$
\begin{equation}
	H_{\alpha\beta}=-p_\mu(M_{\alpha\beta})^\mu_\nu q^\nu=p_\beta\eta_{\alpha\nu}q^\nu- p_\alpha\eta_{\beta\nu}q^\nu  
\end{equation}
satisfying
\begin{equation}
\label{realisation_Lorentz}	\{H_{\alpha\beta},H_{\gamma\xi}\}=\eta_{\beta\gamma}H_{\alpha\xi}-\eta_{\alpha\gamma}H_{\beta\xi}+\eta_{\alpha\xi}H_{\beta\gamma}-\eta_{\beta\xi}H_{\alpha\gamma}\,.
\end{equation} 
The Lagrangian one-form corresponding to the (connected component of) the Lorentz group thus reads
$$L=p_\mu dq^\mu-\frac{1}{2}H_{\alpha\beta}\theta^{\alpha\beta}\,.$$
From our general results, we know that the solutions of the Euler-Lagrange equations associated to $L$ must give us a Hamiltonian action of the Lorentz group on $M$. From the present setup, we clearly expect this action to be nothing but the usual action of the (connected component of the) Lorentz group on the (spacetime) coordinates $q^\nu$ and their momenta $p_\mu$:
\begin{equation}
	\label{action}
	q^\nu\mapsto q{'}^\nu=\Lambda^\nu_\mu q^\mu\,, ~~p_\mu\mapsto p{'}_\mu=\Lambda^\nu_\mu p_\nu\,.
\end{equation}
Let us check that \eqref{non_auto1}-\eqref{non_auto2} indeed produce the desired Hamiltonian action. By design,   \eqref{non_auto3} (equivalently \eqref{Hc}) is satisfied, see \eqref{realisation_Lorentz}. Next, recalling that we chose to parametrise the generators with two indices, \eqref{non_auto1}-\eqref{non_auto2} give the following system of compatible equations
\begin{align}
	\frac{\partial q^\mu}{\partial t^{\sigma\xi}}&=\theta^{\alpha\beta}_{\sigma\xi}\frac{\partial H_{\alpha\beta}}{\partial p_\mu}=-(M_{\alpha\beta})^\mu_\nu \theta^{\alpha\beta}_{\sigma\xi} q^\nu\,,\label{Lorentz1}\\
	\frac{\partial p_\mu}{\partial t^{\sigma\xi}}&=-\theta^{\alpha\beta}_{\sigma\xi}\frac{\partial H_{\alpha\beta}}{\partial q^\mu}=p_\mu(M_{\alpha\beta})^\mu_\nu\theta^{\alpha\beta}_{\sigma\xi}\,.\label{Lorentz2}
\end{align}
In the representation \eqref{representation_L}, a group element is given by $\Lambda=e^{-t^{\alpha\beta}M_{\alpha\beta}}$. Then, \eqref{Lorentz1} is equivalent to 
$$d((\Lambda^{-1})^\mu_\nu q^\nu)=0\,,$$
so $q^\nu=(\Lambda)^\nu_\mu q^\mu_0$ for some ``initial condition'' $q^\mu_0$ at $t^{\alpha\beta}=0$. Thus the solution flow indeed yields the action \eqref{action} of the Lorentz group on $q^\mu$. The reasoning for $p_\mu$ and \eqref{Lorentz2} is similar.

\section{Conclusion and outlooks}\label{sec:ccl}
We introduced an equivalent reformulation of the variational principle at the basis of Lagrangian multiform theory as a variational framework for integrability. Importantly, this replaces the previous two-step formulation with a single {\it univariational} principle, with the effect of setting dependent and independent variables on the same footing. Doing so, we revealed the geometry of Lagrangian one-forms and used it to extend them beyond integrable hierarchies to the realm of non abelian Lie groups. As a consequence, we obtained for the first time the description of Hamiltonian Lie group action as Euler-Lagrange equations derived from a variational principle. 

Our results immediately raise the question of how to reformulate Lagrangian multiforms for field theories along the same lines as done here for finite-dimensional systems. Indeed, the structure of our phase-space one-forms is largely motivated by the Lagrangian one-forms constructed in \cite{CDS,CSV} which are naturally formulated on coadjoint orbits of certain Lie groups characterised by the so-called classical $r$-matrix \cite{STS}. It turns out that the class of Lagrangian multiforms for integrable field theories in $1+1$ dimensions constructed in \cite{CS2,CSV2} possess the same fundamental structure. Therefore, it seems natural to try to extend the present work to the context of these field theories. 

We also believe that our results bring us one step closer to the path integral quantisation of integrable hierarchies based on Lagrangian multiforms. Beyond the need to restore the balance between Hamiltonian and Lagrangian formalisms in integrable systems, this is one of the strongest motivations for Lagrangian multiforms. In the first work \cite{KN} on this topic (see also \cite{KY}), the open question of how to formulate a quantum propagator over paths not only in the degrees of freedom but also in the multi-time is raised as the fundamental problem to overcome. With our approach, this can be done in principle, with the same level of ``rigour'' as the usual Feynman integrals, using our phase-space Lagrangian one-forms and paths $\gamma$ into $M\times G$ in the path integral ``measure''.   

\paragraph{Acknowledgement} 
The authors are very thankful to the anonymous referee for highlighting the connection between their Lagrangian one-form \eqref{Lag_on_G} and the theory of symplectic quotients, which is included in Section \ref{sec:group general}.

\paragraph{Conflict of interest statement}
On behalf of all authors, the corresponding author states that there is no conflict of interest.

\paragraph{Data Availability} This manuscript has no associated data.

\end{document}